\documentclass[%
reprint,
footinbib, 
amsmath,amssymb,
aps,
pre,
]{revtex4-1}

\usepackage{bm}
\usepackage{color}
\usepackage[dvipdfmx]{graphicx}
\usepackage{comment}
\begin{document}
	\title{Aversion of face-to-face situation of
		pedestrians eases crowding condition}
	
	\author{Sho Yajima}
	\affiliation{%
		Department of Applied Physics, Faculty of Science Division I, Tokyo University of Science, 6-3-1 Nijuku, Katsushika, Tokyo, 125-8585, Japan}
	\author{Kiwamu Yoshii}
	\affiliation{%
		Earthquake Research Institute, The University of Tokyo, 1-1-1, Yayoi, Bunkyo-ku, Tokyo 113-0032, Japan}
	\author{Yutaka Sumino}
	\email{ysumino@rs.tus.ac.jp}
	\affiliation{%
		Department of Applied Physics, Faculty of Science Division I, Tokyo University of Science, 6-3-1 Nijuku, Katsushika, Tokyo, 125-8585, Japan}
	\affiliation{Water Frontier Science \& Technology Research Center and I$^2$ Plus, Research Institute for Science \& Technology, Tokyo University of Science, 6-3-1 Nijuku, Katsushika, Tokyo, 125-8585, Japan
	}
	\date{\today}
\begin{abstract}
	We conducted numerical simulation for a crowd of pedestrians. Each pedestrian, modeled with three circles, has a shape whose long axis is perpendicular to the anteroposterior axis, and is designed to move fixed destination. The pedestrians have friction at the surface and soft repulsion. In this study, we newly introduced an active rotation which captures psychological effect to evade face-to-face situation. The numerical simulation revealed that active rotation induces fluidization of system leading to higher flux of pedestrian. We further confirmed that this fluidization is due to fragmentation of force chain induced by the active rotation.
\end{abstract}

\maketitle

\section{Introduction}
Active matter is a systems composed of many elements that transduces energy into motion locally. Such active matters are novel type of nonequilibrium system, showing rich dynamical pattern formation. Bird flocks, fish schools, and bacterial colonies are well known examples for active matters \cite{ramaswamy2010mechanics, vicsek2012collective,marchetti2013hydrodynamics, bechinger2016active}.
The population of human is also considered as active matter and show a variety of pattern formation \cite{helbing2005self, ikeda2012instabilities, Poncet2017}.

We here make a model for crowds of pedestrian in high density. 
It is known that a crowd of pedestrian in high density leads to fatal accidents\cite{helbing2000simulating, frank2011room, karamouzas2014universal}. Such critical situation is due to the formation of force chain within the population, and it leads to stress concentration to small number of pedestrians.
Many types of numerical simulation is conducted to understand the behavior of pedestrians in high density\cite{helbing1999optimal, castellano2009statistical, ikeda2017lane}.
In such numerical simulation, a pedestrian can be modeled with active granular particle that follows Newton's equation of motion, where each individual walks to their fixed destinations~\cite{sfm, counterflow}.

A pedestrian has anisotropy in function and shape: it has frontal part directing their motion, and is wider in the direction perpendicular to the anteroposterior axis \cite{thompson1995computer}.
In this study, we newly adopted an active rotation induced by psychological effect. Front direction is preferred direction to move. Interestingly, however, people psychologically avert face-to-face situation each other. We adopt this effect by the active rotation. Our numerical simulation reveals that the coupling between anisotropic shape and active rotation eases crowded condition for pedestrians. 


\begin{figure}[b]
\includegraphics{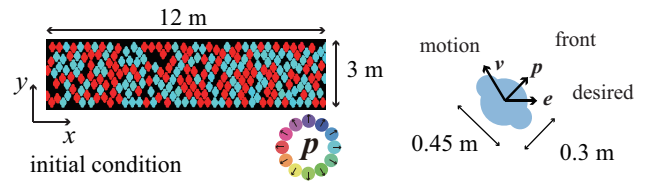}
\caption{Schematic illustration of a pedestrian model. Initial condition of numerical simulation. Pedestrians of fixed ratio, $R$, walk positive $x$ direction, whereas the other walk negative $x$ direction. Such desired direction is denoted by  $\bm{e}_{i}$. Each pedestrian has shape composed of three circles, which has the front direction $\bm{p}_i$. Central one is twice as large as the other ones, The centers of the side ones situate on the arc of the central one, and perpendicular to $\bm{p}_{i}$. Each pedestrian has also the velocity vector $\bm{v}_{i}$. The color represents the direction of $\bm{p}$.}
\label{figure:1}
\end{figure}

\section{Simulation setup}
As shown in Fig.\ref{figure:1}(a), we consider rectangle system, i.e $L_x =$~12 m and $L_y =$~3 m, and impose the periodic boundary condition on $x$ direction. Each pedestrian has desired direction $\bm{e}_{i} = (\pm 1, 0)$, front direction $\bm{p}_i = (\cos\theta_i, \sin\theta_i)$ ($\theta_{i} \in [-\pi,~\pi)$), and velocity $\bm{v}_i$.
In this study, we model a pedestrian $i$ with central circle whose radius is $\ell_i$, and side circles whose radius is $\ell_i/2$. $\ell_i$ is fixed for each pedestrian and mean of $\ell_i$ is 0.15 m, with $\pm$ 5 \% homogeneous random  distribution around the mean. 
The position of center is denoted by $\bm{r}_{i}=({x}_{i},{y}_{i})$.
The centers of the side ones situate on the arc of the central one, and perpendicular to $\bm{p}_{i}$.
The position of right circle is, then, 
$\bm{r}_{i}^{r}=({x}_{i}^{r},{y}_{i}^{r})=({x}_{i}+\frac{\ell_i}{2}\sin{\theta}_{i},{y}_{i}-\frac{\ell_i}{2}\cos{\theta}_{i})$, and that of left circle is
$\bm{r}_{i}^{l}=({x}_{i}^{l},{y}_{i}^{l})=({x}_{i}-\frac{\ell_i}{2}\sin{\theta}_{i},{y}_{i}+\frac{\ell_i}{2}\cos{\theta}_{i})$. We can define
the velocity of central circle is $\bm{v}_{i}=\dot{\bm{r}}_{i}=(v_{x,i},v_{y, i})$, and those of left circle and right circle are 
$\bm{v}_{i}^{r}=(v_{x,i}^{r},v_{y, i}^{r})=({v}_{x, i}+\frac{\ell_i{\omega}_{i}}{2}\cos{\theta}_{i},{v}_{y, i}+\frac{\ell_i{\omega}_{i}}{2}\sin{\theta}_{i})$ and $\bm{v}_{i}^{l}=({v}_{x, i}^{l},{v}_{y, i}^{l})=({v}_{x,i}-\frac{\ell_i{\omega}_{i}}{2}\cos{\theta}_{i},{v}_{y, i}-\frac{\ell_i{\omega}_{i}}{2}\sin{\theta}_{i})$, respectively. $\omega_i$ represents the angular velocity of rotation.

The equation of motion for $i$-th pedestrian is given by
\begin{align}\label{EOM}
{m}_{i}\frac{d\bm{v}_{i}}{dt} =  m_{i}\frac{v_{0}\bm{e}_{i}-\bm{v}_{i}}{\tau}+\sum_{j\neq i}\bm{f}_{ij}+\sum_{W}\bm{f}_{iW}+\bm{F}^{b}_{i}, 
\end{align}and
\begin{align}\label{RotEOM}
I_i\dot{\omega}_{i}=&\sum_{j\neq i}({N}^{\mathrm{phy}}_{ij}+{N}^{\mathrm{psy}}_{ij})+\sum_{W} {N}^{\mathrm{phy}}_{iW} \nonumber \\
&-\eta_{1}\cos^{-1}(\bm{p}_i \cdot\bm{e}_i) -\eta_{2}{\omega}_{i}+\xi(t).
\end{align}
Eq.~\eqref{EOM} represents the dynamics of the center of the mass, and Eq.~\eqref{RotEOM} represents the dynamics of the rotation.

In Eq.~\eqref{EOM},
$m_i$ is the mass of the pedestrian, $v_{0}$ is the desired speed of pedestrian, $\tau$ is the relaxation time required by a pedestrian to reach the desired speed, unit vector $\bm{e}_{i}$ is the desired direction of motion, which is either $(1, 0)$ or $(-1,0)$. $\bm{f}_{ij}$ represents contact force acting on the $i$-th pedestrian from $j$-th one, $\bm{f}_{iW}$ acting on the $i$-th pedestrian from wall particles.
We also introduce back force, direction dependent resistance, $\bm{F}^{b}_{i}$ as,
$\bm{F}^{b}_{i}=A\Theta(-\bm{e}_{i}\cdot\bm{v}_{i})(m_{i}v_{0}/\tau)[\bm{e}_{i}-(\bm{e}_{i}\times\bm{e}_{z})\tanh{\psi}]$.~Here, $\psi=-\cos^{-1} \left(\bm{v}_i \cdot \bm{p}_i/|\bm{v}_i| \right)$ and $A$ is the intensity factor. We also used Heaviside's step
function $\Theta(x)$ satisfying $\Theta(x) = 1$ for $x \geq 0$ and $\Theta(x) = 0$ otherwise. 
Unit vector $\bm{e}_{z}$ is the direction of the symmetry axis. This back force is also introduced in the prior study\cite{counterflow}.
The angular velocity $\omega_{i}=\dot{\theta}_i$ changes by the following rotational equation of motion given by Eq.~\ref{RotEOM}.

In Eq.~\ref{RotEOM}, the third term in the right hand side represents the tendency of a pedestrian to align themselves to the desired direction, ($\theta_i = 0$ or $\pi$, depending on $\bm{e}_i$). The fourth term is rotational friction, and the fifth term represents a Gaussian white noise which is defined by $\langle \xi(t) \rangle=0$ and $\langle \xi(t) \xi(t') \rangle= \sigma^2 \delta(t-t')$. In the first and the second terms, we introduced two types of torque. One is the physical torque $N^{\mathrm{phy}}_{ij}$ and $N^{\mathrm{phy}}_{iW}$ caused by contact to another pedestrians and walls. The other one is psychological torque $N^{\mathrm{psy}}_{ij}$ to avoid being face-to-face situation to another pedestrians, and this term is newly introduced in our study.

The details of physical force is based on the prior research~\cite{counterflow}.
A pedestrian is, as mentioned, made of combination of circles. When two pedestrian touches each other, the contact is made in a form of contact point. Here, we incorporate repulsion due to excluding volume and friction. The surface is assumed to indent each other whose normal distance is $\delta$, and has relative speed $\bm{v}_{\mathrm{rel}}=v^n_{\mathrm{rel}}\bm{n}+v^t_{\mathrm{rel}}\bm{t}$. To incorporate the coulomb friction, we define $\Xi$ from the relative tangential velocity as 
\begin{equation}
\Xi(t)=\int^t_{t' = ts} v_{\mathrm{rel}}^{t}(t') dt',
\end{equation}
where $t_s$ is the start time when two surface contact.
Then the force acting at the contact point is modeled as
\begin{equation}
\bm{f}={F}^{n}\bm{n}+{F}^{t}\bm{t},
\end{equation}
where $\bm{n}$ and $\bm{t}$ are outward normal and tangential to the contact surface.
Here, ${F}^{n}\bm{n}$ is soft core repulsion with viscous damping effect described by 
\begin{align}
{F}^{n}=-{k}^{n}\delta-{\gamma}^{n}v^n_{\mathrm{rel}}.
\end{align}
${F}^{t}\bm{t}$ describes coulomb friction~\cite{xi} as
\begin{align}
{F}^{t}=
\left\{
\begin{array}{cc}
-{k}^{t}\Xi-{\gamma}^{t}v_{\mathrm{rel}}^{t} & (|{F}^{t}| \leq -\mu F^{n})\\
-\mu F^{n} & (|{F}^{t}| > -\mu F^{n})
\end{array}
\right.   
\end{align}

\begin{figure}\label{pedestrians}
	\begin{center}
		\includegraphics{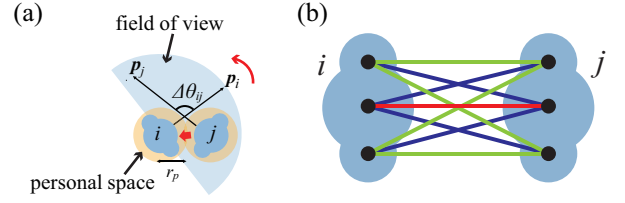}
		\caption{ (a)~Active rotation of pedestrian reflecting psychological effect. When other pedestrians are within a pedestrians field of view, and they direct face to face, the pedestrian will turn itself.(b)~The interaction between pedestrian $i$ and $j$. Contact can be possible between central and central (red), side to central (blue), and side to side (green) circles. }
		\label{figure:2}
	\end{center}
\end{figure}

The physical force $\bm{f}_{ij}$ and torque $N^{\mathrm{phy}}_{ij}$ on pedestrian $i$ from $j$ consist of the contact forces between central and central, side to central, and side to side circles represented as red, blue, and green lines in Fig \ref{figure:2}(b).

Contact force between central circle on $i$ from $j$, $\bm{f}_{ij}^{cc}$ is calculated with following way.
The distance between the central circles of $i$ and $j$, is given by $d=|\bm{r}_{i}-\bm{r}_{j}|$.
Then, we have $\bm{n}=(n_x,n_y)=(\bm{r}_{j}-\bm{r}_{i})/d$, and  $\bm{t}=(-n_y,n_x)$, and
\begin{align}
\delta =
\left\{
\begin{array}{cc}
0   &  (d\geqq{\ell}_{i}+{\ell}_{j})\\
{\ell}_{i}+{\ell}_{j}-d   & (d<{\ell}_{i}+{\ell}_{j})
\end{array}
\right.
\end{align}
The relative velocity is
\begin{equation}
v_{\mathrm{rel}}^{n}=(\bm{v}_{i}-\bm{v}_{j})\cdot\bm{n},
\end{equation}
and
\begin{equation}
v_{\mathrm{rel}}^{t}=(\bm{v}_{i}-\bm{v}_{j})\cdot\bm{t}+{\ell}_{i}{\omega}_{i}+{\ell}_{j}{\omega}_{j}.
\end{equation}
Note that the effect of rotation is also included by ${\ell}_{i}{\omega}_{i}$ and ${\ell}_{j}{\omega}_{j}$.

\begin{figure*}[t]
	\includegraphics{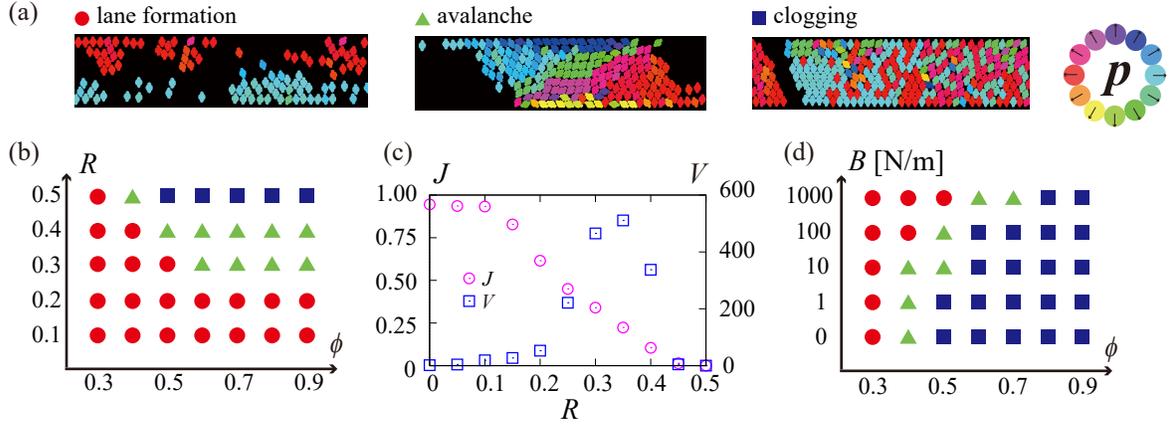}
	\caption{Typical behavior of simulation and phase diagrams. Each symbol represents respectively lane formation (\textcolor{red}{$\bullet$}), avalanche (\textcolor{green}{$\blacktriangle$}) and clogging(\textcolor{blue}{\tiny$\blacksquare$}). (a) Snapshots of three characteristic regimes observed with our model. The color represents the direction $\bm{p}$.~(b)Phase diagram for the present system with controlled parameters as packing area fraction $\phi$ and the population ratio $R$.~(c)We show population ratio against the average flow rate $J$ and variance of flow rate $V$, when $\phi = 0.7$.~(d)~Phase diagram for the present system with parameters as $\phi$ and a psychological effect $B$.}
	\label{figure:3}
\end{figure*}

The contact force on the central circle of $i$ from the side circle $j$ is calculated in the following manner. Here we exemplify the contact force $\bm{f}_{ij}^{cr}$, the force on central circle of $j$ from the right circle of $i$.
Now, we have $d=|\bm{r}_{i}-\bm{r}_{j}^{r}|$, 
$\bm{n}=(n_x,n_y)=(\bm{r}_{j}^{r}-\bm{r}_{i})/d$, and $\bm{t}=(-n_y,n_x)$.
\begin{align}
\delta =
\left\{
\begin{array}{cc}
0   &  (d\geqq{\ell}_{i}+\ell_j/2)\\
{\ell}_{i}+{\ell}_{j}/2-d   & (d<{\ell}_{i}+{\ell}_{j}/2)
\end{array}
\right.
\end{align}
The relative velocity is now,
\begin{equation}
v_{\mathrm{rel}}^{n}=(\bm{v}_{i}-\bm{v}_{j})\cdot\bm{n},
\end{equation}
and
\begin{equation}
v_{\mathrm{rel}}^{t}=(\bm{v}_{i}^{l}-\bm{v}_{j}^{r})\cdot\bm{t}+{\ell}_{i}{\omega}_{i}.
\end{equation}

The contact force on the side circle of $i$ from the side circle $j$ is calculated in the following manner. Here we exemplify the contact force $\bm{f}_{ij}^{lr}$, the force on the left circle of $j$ from the right circle of $i$.
Now, we have $d=|\bm{r}_{i}^{l}-\bm{r}_{j}^{r}|$, 
$\bm{n}=(n_x,n_y)=(\bm{r}_{j}^{r}-\bm{r}_{i}^{l})/d$, and $\bm{t}=(-n_y,n_x)$.
\begin{align}
\delta =
\left\{
\begin{array}{cc}
0   &  (d\geqq{\ell}_{i}/2+\ell_j/2)\\
{\ell}_{i}/2+{\ell}_{j}/2-d   & (d<{\ell}_{i}/2+{\ell}_{j}/2)
\end{array}
\right.
\end{align}

The relative velocity is now,
\begin{equation}
v_{\mathrm{rel}}^{n}=(\bm{v}_{i}^{l}-\bm{v}_{j}^{r})\cdot\bm{n},
\end{equation}
and
\begin{equation}
v_{\mathrm{rel}}^{t}=(\bm{v}_{i}^{l}-\bm{v}_{j}^{r})\cdot\bm{t}.
\end{equation}

Finally we have, 
\begin{equation}
\bm{f}_{ij}=\sum_{\alpha}\bm{f}_{ij}^{\alpha}.
\end{equation}
Here $\alpha$ is 9 possible combination of $c$, $l$ and $r$.
We also have
\begin{equation}
{N}^{\mathrm{phy}}_{ij}=\sum_{\alpha}(\bm{R}_{\alpha}-\bm{r}_{i})\times\bm{f}_{ij}^{\alpha},
\end{equation}
where $\bm{R}_{\alpha}$ represents the position of each contact point.

The interactions with walls $\bm{f}_{iW}$ and $N^{\mathrm{phy}}_{iW}$ are also calculated in the manner of interaction between pedestrian. Wall is described by the continuously arranged particle with fixed size $r_W$ and no rotation. The particle was set whose distance was $2r_W$. 

As for $N^{\mathrm{psy}}_{ij}$, the visual field region of $i$-th pedestrian is defined as semicircle ($\theta_{i} -\pi,~\theta_{i} + \pi$) region whose radius is $1$ m (Fig \ref{figure:2}(a)). When another pedestrians denoted by $j$ comes into this region, and the $j$-th pedestrian facing to $i$-th pedestrian ($\Delta \theta_{ij}=\theta_j-\theta_i < \pi/2$), $i$-th pedestrian experiences psychological aversion to make itself turns. This effect is adopted as $N^{\mathrm{phy}}_{ij}$:
\begin{align}
{N}^{\mathrm{psy}}_{ij}=\bm{e}_z\cdot(\bm{f}^{\mathrm{psy}}_{ij}\times{\bm{p}_{i}}).
\end{align}
Here, $\bm{f}^{\mathrm{psy}}$ is psychological force affecting only rotation, and definced as
\begin{align}
\bm{f}^{\mathrm{psy}}_{ij}=
\begin{cases}
B(2r_p -|\bm{r}_{ji}|)\dfrac{\bm{r}_{ji}}{|\bm{r}_{ji}|}&(|\bm{r}_{ji}| < 2r_p\cap|\Delta \theta_{ij}| > \frac{\pi}{2}),\\ 
0 &(\mathrm{otherwise}),
\end{cases}
\end{align}
Here, we define relative position vector $\bm{r}_j-\bm{r}_i$ as $\bm{r}_{ji}$.

We mainly used parameters according to prior study\cite{counterflow}. 
As mentioned the mean of $\ell_{i}=0.15$ m, and distributed in 0.1425-0.1575 m with homogeneous random distribution. $v_{0}$ is randomly distributed in 0.8-1.2 m/s. $m_{i}$ is also randomly distributed in 40-70~kg. 
Other parameters are as follows: 
$L_x=12$ m, $L_y=3$ m, $r_{W}=0.1$ m, $\tau=0.67$ s, $I=1.0$ kg$\cdot$m$^2$, ${k}^{n}=1.0\times10^5$ N/m, ${k}^{t}=1.0\times10^4$ N/m, ${\gamma}^{n}=1.0\times10^1$ 1/s, ${\gamma}^{t}=1.0\times10^1$ 1/s, $\mu=0.4$, $\eta_{1}=15$ N$\cdot$m, $\eta_{2}=20$ N$\cdot$m$\cdot$ s, $r_{p}=0.25$ m, and $\sigma =1$ N$\cdot$m$\cdot$ s$^{1/2}$

We integrated the equation with the Euler-Maruyama scheme using $\Delta t=0.001$ s. Initially, pedestrians are randomly distributed in the system. The interaction terms, the size of particles are increase gradual manner from $t$ = 0 to 10 s to avoid spatial overlap. The system is settled down up to $t=$ 20 s. Then, after these terms fully introduced, rest of terms were introduced to begin numerical simulation of Eq. \eqref{EOM} and \eqref{RotEOM}.

\begin{figure}[t]
	\includegraphics{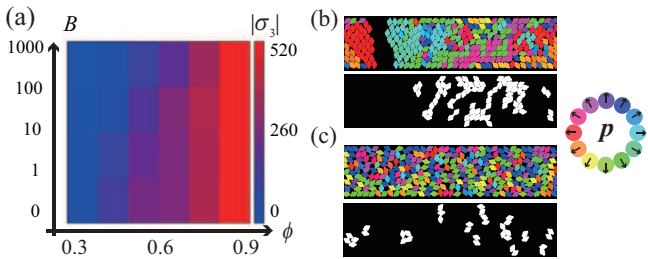}
	\caption{(a) Averaged size of compressive stress $|\sigma_3|$ with various parameters $\phi$ and $B$ with $R=0.5$. (b, c)Typical snapshots in denser system ($\phi=0.8$) and force chain network.~(b)~When $B=0$, i.e psychological effect is weak, long and broad force chain network observed. ~(c)~When $B=1000$, i.e psychological effect is strong, short and localized force chain network observed. For the effect of active rotation, the force chain structures disappear.}
	\label{figure:4}
\end{figure}

\section{Result and Discussion}
The results of the our model is exemplified in Fig.~\ref{figure:3}. In figure~\ref{figure:3} (a)-(c), we show the behavior when $B=0$. This situation is almost the same as prior study. Figure~\ref{figure:3} (a) shows typically observed behavior of current model: lane formation, avalanche, and clogging. The phase diagram of current system with parameters: the packing  ratio $\phi$ and the population ratio $R$ is shown in Fig.~\ref{figure:3} (b). 
We measured the number of pedestrian $N_{\mathrm{th}}$ crossing boundary at the left and the right in $T=$~10 s. Here the ideal number of pedestrian crossing boundary without collision $N_{\mathrm{id}}$ in $T$ s is estimated by $N_{\mathrm{id}}=v_{0} T N_{\mathrm{tot}}/L_x $, where $N_{\mathrm{tot}}$ is total number of pedestrians. We define flow rate $j=N_{\mathrm{th}}/N_{\mathrm{tot}}$, and its average and variance is defined by $J=\langle j \rangle$ and $V = \langle (j-J)^2 \rangle$.
$J$ and $V$ for $\phi=0.7$ is plotted in Fig.~\ref{figure:3} (c). Flow rate is defined by the number ratio of particle crossed boundary within When $\phi$ and $R$ is small, pedestrians flow smoothly, with high $J$ and small $V$ while making lane structure. At high $\phi$ and $R\sim0.5$, average flow rate $J$ becomes 0, and they shows clogged structure. In between these parameters, the fluctuation of flow rate $V$, where clogging and flowing alternated stochastic manner: avalanche. These results are consistent with the prior study~\cite{counterflow}.

Active rotation due to psychological effect increase fluidity of the system as shown in Fig.~\ref{figure:3}(d).
When $B$ is increased, even clogging situation becomes smoothly flowing lane formation phase. This result is counter intuitive, as increase of the psychological repulsion increase effective occupation area, and should enhance clogging.

We analyzed the physical stress in detail to grasp the essential mechanism of fluidization due to the effect of $B$. We obtained minimum principal stress for each pedestrian $\sigma_3$, which is essentially compressive stress. The averaged value of the size $|\sigma_3|$ is plotted in Fig.~\ref{figure:4}(a). As seen, $|\sigma_3|$ decreases with the increase in $B$, when the system become fluidized.
The snapshots as well as force chain structure obtained to visualize the effect of $B$. The force chain is reconstructed based on the algorithm proposed by Peters et al.\cite{forcechain}.
Here we show the typical snapshot at $\phi=0.8$ (Fig.~\ref{figure:4}(b, c)) where system stays as clogged even with $B=1000$. In Fig.~\ref{figure:4}(b), the data with $B=0$ is shown. Large void space is noted, with long force chain. By contrast, with $B=1000$, void space is now absent and pedestrians are homogeneously distributed. The force chain is now localized, and does not span whole system. It is known that the bridging of force chains between the boundary fix particles in the chains.
We confirmed that the increase in $B$ leads larger effective occupation area as expected. However, the active rotation $B$  induces fragmentation of force chain, finally results in the decrease of average compression force $|\sigma_3|$. We may also recognize such fluidization is due to the flow induced by the collective effect of active rotor\cite{Nguyen2014}. 

\section{Conclusion}
In this study, we considered a model of 
pedestrians with anisotropic shape and surface friction, as well as the active rotation due to psychological effect. A pedestrian is modeled by the combination of three circles walking towards two different direction. A pedestrian experiences physical force and torque due to excluding volume effect and coulomb friction. We newly introduced active torque, modeling psychological effect to evade face-to-face situation. We confirmed that the active rotation increase the effective occupation area for each pedestrian, but it also fragments force chain structures. Overall, the system is fluidized by the effect of the active rotation.

\subsection*{Acknowledgment}
This work was supported by JSPS KAKENHI Grant Nos. JP16K13866, JP16H06478 and JP19H05403. This work was also partially supported by a JSPS Bilateral Joint Research Program between Japan and the Polish Academy of Sciences. ``Spatio-temporal patterns of elements driven by self-generated, geometrically constrained flows", and the Cooperative Research of the ``Network Joint Research Center for Materials and Devices" with Hokkaido University (No. 20181048).

\end{document}